\documentclass[sigconf]{acmart}

\usepackage{booktabs} 
\usepackage{graphicx}
\graphicspath{ {images/} }

\setcopyright{rightsretained}

\acmDOI{10.475/123_4}

\acmISBN{123-4567-24-567/08/06}

\begin{document}

\copyrightyear{2017}
\acmYear{2017}
\setcopyright{acmlicensed}
\acmConference{PEARC17}{July 09-13, 2017}{New Orleans, LA, USA}\acmPrice{15.00}\acmDOI{10.1145/3093338.3093370}
\acmISBN{978-1-4503-5272-7/17/07}

\title{Portable Learning Environments for Hands-On Computational Instruction}
\subtitle{Using Container- and Cloud-Based Technology to Teach Data Science}

\author{Chris Holdgraf}
\affiliation{%
  \institution{Berkeley Institute for Data Science, University Of California, Berkeley}
  \streetaddress{190 Doe Library}
  \city{Berkeley}
  \state{CA}
  \postcode{94720}
}
\email{choldgraf@berkeley.edu}

\author{Aaron Culich}
\affiliation{%
  \institution{Research IT, University Of California, Berkeley}
  \streetaddress{250-26 Warren Hall}
  \city{Berkeley}
  \state{CA}
  \postcode{94720}
}
\email{aculich@berkeley.edu}

\author{Ariel Rokem}
\affiliation{%
  \institution{University of Washington e{S}cience Institute}
  \streetaddress{The WRF Data Science Studio}
  \city{Seattle}
  \state{WA}
  \postcode{98105}
}
\email{arokem@uw.edu}

\author{Fatma Deniz}
\affiliation{%
  \institution{Berkeley Institute for Data Science, University Of California, Berkeley}
  \streetaddress{190 Doe Library}
  \city{Berkeley}
  \state{CA}
  \postcode{94720}
}
\email{fatma@berkeley.edu}

\author{Maryana Alegro}
\affiliation{%
  \institution{University of California, San Francisco}
  \streetaddress{Mission Bay}
  \city{San Francisco}
  \state{CA}
  \postcode{94158}
}
\email{maryana.alegro@ucsf.edu}

\author{Dani Ushizima}
\affiliation{%
  \institution{Lawrence Berkeley National Lab and Berkeley Institute for Data Science, University Of California, Berkeley}
  \streetaddress{}
  \city{}
  \state{}
  \postcode{}
}
\email{dushizima@lbl.gov}

\renewcommand{\shortauthors}{Holdgraf et al.}

\begin{abstract}

There is an increasing interest in learning outside of the traditional classroom
setting. This is especially true for topics covering computational tools and data science,
as both are challenging to incorporate in the standard curriculum.
These atypical learning environments offer new opportunities for
teaching, particularly when it comes to combining conceptual
knowledge with hands-on experience/expertise with methods and skills.
Advances in cloud computing and containerized environments
provide an attractive opportunity to improve the efficiency and ease with which students
can learn. This manuscript details recent advances towards using
commonly-available cloud computing services and advanced cyberinfrastructure
support for improving the learning experience in bootcamp-style events. We cover
the benefits (and challenges) of using a server hosted remotely instead of
relying on student laptops, discuss the technology that was used in order to
make this possible, and give suggestions for how others could implement and
improve upon this model for pedagogy and reproducibility.

\end{abstract}




\maketitle

\section{Introduction}

Science has recently seen a large growth in the use of
computationally intensive and data-centric methods. Researchers
increasingly use open-source programming languages such as R or Python, and utilize
complex algorithms in applied statistics and machine learning in order to
perform their research \cite{momcheva2015astro}.

With an increased focus on computational methods
comes new challenges in teaching these techniques and new approaches
toward sharing knowledge with fellow scientists.
A rapidly growing approach to scientific training involves learning
outside of the traditional semester-long classroom setting.
This is particularly true
for teaching computational and data analytic techniques,
as instructors must teach both conceptual and methodological information
simultaneously. A common way to teach these skills is a short,
time-bounded learning event, such as a bootcamp or workshop
\cite{wilson2016software}. These  courses attempt to compress several
topics into an intensive learning session that is usually held over one or
several days.

As these new learning models are adopted, it opens
opportunities for developing new technology and models for pedagogy that are
focused on "hands-on" learning. Here we describe
recent work in utilizing cloud-based infrastructure to enhance this learning
experience, and to streamline the ability of instructors to
teach material that focuses on data analytic techniques. Our approach
utilizes recent advances in cloud- and cluster- based technology, such
as container technology (e.g., Docker) and the
Jupyter framework for computing.

This article is a case-study covering our recent experience implementing this approach, using
advanced cyberinfrastructure to teach a multi-institutional day-long bootcamp in machine learning
hosted at the University of California in San Francisco (UCSF).
We describe the technical tool-chain and the processes that were used in
designing this course. This
unique design includes hosting all course materials online and providing
an interactive, online environment where students can run code via the cloud
without requiring them to download anything onto their machine.
It is a snapshot of the current state
of technology and practice around these ideas, and is likely to evolve rapidly
as both the tools and our knowledge of their use in pedagogy improves.

Importantly, this software stack only utilizes tools that are
open-source and freely available to
the community, and uses low-cost computing services that are available
to instructors and institutions in the United States. In addition, the
use of these technologies does not require that the entire
team of instructors become fully skilled at using them. It is
usually enough for a single team member to understand, set up, and manage these resources.
The container technology (Docker) and cloud
computing resources (XSEDE) we use are widely available platforms.
However, this teaching environment can be similarly applied using
different container or cloud computing resources.
We will discuss the challenges in implementing this course
setup effectively, and discuss its merits and drawbacks.

\subsection{The bootcamp model of pedagogy}

For the purposes of this paper we define bootcamps and workshops within
the same category of time-bounded events. These are relatively short-term
learning sessions in which a group of students moves rapidly through a
collection of training material, generally with the guidance of one
or many instructors. Time-bounded workshops often follow the same
formula, roughly described here.

First, instructors develop materials on their own computers, sharing them with
participants (e.g., as a public Github repository). Often, instructors will use
file formats, such as Jupyter notebooks, that
interleave code, text describing the
data and computations, results, and illustrations
\cite{kluyver2016jupyter}. In the days preceding the event, organizers send
instructions to participants, such as how to download the materials and their
dependencies or how to configure these dependencies on their
laptop computers. On the day of the workshop, instructors assume that
students have already followed these instructions successfully, or hold
mini ``install-fest'' sessions that assist students in
getting their environments set up. The course itself emphasizes hands-on
learning, and students interact with course material on their own
machines as the instructors teach.

This kind of time-bounded bootcamp offers many advantages for learning over
longer courses. For example, they allow the students to focus
entirely on one topic for an extended period of time. This can be particularly
useful for material that demands a ``deeper dive'' and intensive, hands-on
work. It is also particularly useful for topics that jointly
cover both conceptual material and more ``hands-on'' tasks, because the
condensed time leaves more room for experimentation, discussion, and active
learning \citep{Bransford2000-lu, Papert1980-fh}. In addition, due to the
interactive setup, students are more likely to work collaboratively and learn
from each other.

However, there are still many challenges associated with this class structure.
Students interact with material on their own, and their learning experience is
heavily dependent on the ability of each student to get started in the first
place. Because the instructional materials are developed on instructor
computers, differences between the instructor and student computers (e.g., memory
available, operating system, etc.) can have hard-to-predict consequences such
as slow execution (e.g., due to sub-par student hardware) or a failure to
execute code (e.g. because of missing software dependencies on
students' computers). These types of courses are often relatively short,
and small delays result in a significant loss of instructional time.

A solution to mitigate many of these challenges is to offload the issue of
student-specific hardware onto a shared cloud computing platform. This approach
standardizes the experience of each student by allowing them access to a single
online resource for the duration of the class, without requiring any new
software or data to be downloaded on student computers. Students can work
on course material simultaneously with the instructor and can
experiment with multiple solutions on their cloud copy of the material,
creating a more interactive teaching and learning environment. Next we describe
an implementation of this model at a day-long workshop hosted at the
University of California, San Francisco (UCSF).

\section{Results}

\subsection{Course overview and development}

The workshop focused on the analysis and interpretation of neuroimaging data,
ranging from whole-brain functional magnetic resonance imaging (fMRI) to
single-cell microscopy. Instructors were distributed across three institutions,
and attendees were mostly graduate students and post-docs that had minimal
background in data analysis and image processing. The instructors took the
audience through detailed hands-on data analysis pipelines that harness open
source software for neural image processing (\cite{van2014scikit}, \cite{Garyfallidis2014FrontNeuroinf}).
They also introduced the participants to
machine learning techniques (e.g., deep learning methods using Caffe
\cite{jia2014caffe} and Tensorflow \cite{abadi2016tensorflow} as well an
introduction to scikit-learn \cite{Pedregosa2012-dm}).

All course materials were hosted via a shared online computing platform.
Students accessed this platform via web browsers on their own computers.
This made it possible to standardize the computing environment across both
students and instructors, and minimized the effort needed to get students
started with the material. In order to make course materials accessible in a
live, online computing environment, the following considerations were taken:

\begin{enumerate}

\item {\bf Software and computing environment}: The development of the course material was
performed in Docker containers. This ensured that all course materials could
run on any computing architecture that could host such a Docker image.
First, a Docker image
\cite{merkel2014docker} was created with the computational restrictions and
computing environment that would match what would be available in the cloud
platform that was used. Instructors used a shared instance of this Docker
image while they developed sections of the tutorial independently. They used a
shared Github repository to host each section of the bootcamp, and pushed their
materials as separate folders in this repository. As new packages were needed in
order to cover particular topics, the base Docker image was updated with added
dependencies. As a result, by the end of course development there was a single
Docker image with all software tools needed to complete the course.

\item {\bf Data management}: Many bootcamp-style events require data to use in
the materials. For realistic research examples, the data can be cumbersome to
download and modify (for example, if many students attempt to download the
same file at once during class). This course covered a wide range of datasets
including publicly-available fMRI data
as well as images of cells collected from the human brain. Instructors used a
script that fetched this data from online repositories, and then stored it in
a common data folder on the user's computer. The Docker images were then
configured such that this script would be run as soon as a new image was
instantiated, ensuring that all data was pre-loaded onto the user's
filesystem upon launch. Because students had access to their
own computing environments, this data (and any modifications of it) persisted
over time and between computing sessions.

\item {\bf Environment distribution}: In the
day before the class, the course Docker image was deployed to several Virtual Machine (VM)
instances being run on the Jetstream cloud computing platform
\cite{Stewart2015Jetstream}, available through the Extreme Science and
Engineering Discovery Environment (XSEDE) \cite{Towns2014XSEDE}  (see section
\ref{sec:methods}). We chose this environment due to its availability at no cost to many
university campuses, though it is possible to use any cloud provider that
provides Docker support such as Google Cloud, Microsoft Azure, or even
bare-metal hardware running a Linux server.

A single IP address was generated for each instance - one
for each student - and was sent out the morning of the event. Students only
needed to click on their respective IP address, and they were instantly taken to
a live Jupyter notebook contained within their XSEDE Jetstream virtual
machine. This notebook had all course materials, as well as access to the
data, software packages, and live Python kernels that were needed to execute the
course materials. In addition, this environment persisted several days after the
class was finished in order to allow students to continue interacting with the
material, or to port their work onto their personal computers.

\end{enumerate}

\begin{figure}[h]
\centering
\includegraphics[width=0.5\textwidth]{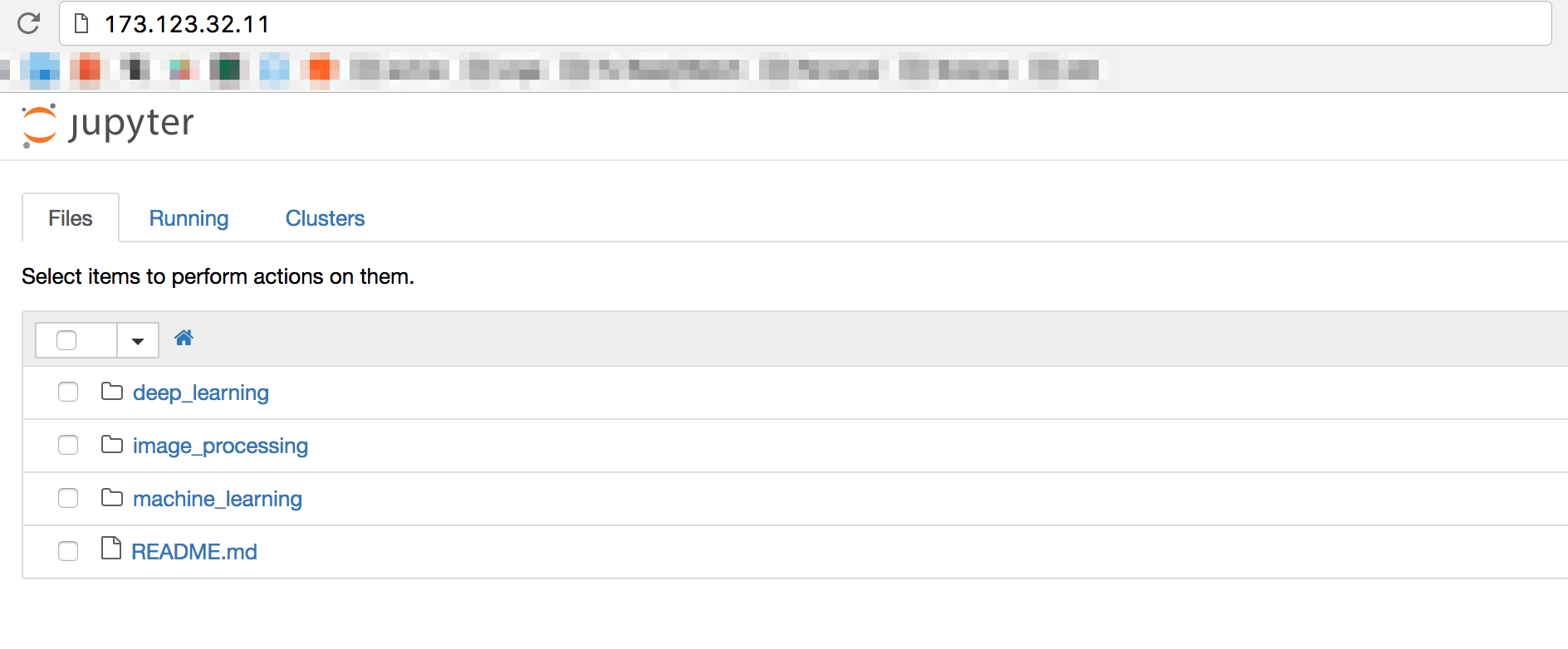}
\caption{Screenshot of a student notebook instance. Students were instructed to
         navigate to an IP address, and the resulting Jupyter notebook was
         displayed with all course materials inside.}
\end{figure}

\subsection{Comparisons with a traditional bootcamp setup}

This section covers some of the main benefits and drawbacks of the cloud
computing approach towards bootcamp-style events. First we will cover primary
benefits of a 'traditional' bootcamp event (one in which participants are asked
to install the software on their own computers). Next we will discuss drawbacks
and how these are addressed by using cloud infrastructure. Finally we will
discuss new challenges that were introduced and how they might be addressed in
the future.

\subsubsection{Benefits of a traditional bootcamp approach}

The challenge of asking students to do all work on their own machines has one
primary benefit: they are working in the computing environment that they
likely use on a daily basis. While it is common for frustrations to pop up
during installation and execution of course materials, these are also
common in every-day practice, and enable incidental learning
of the skills needed to know how to solve computational problems in general.
In addition, by doing all course computation on
their own computer, there is no need to adapt to a completely new computing
approach (beyond learning about new packages, languages, etc). Finally, after
the course is over there is no need to migrate course materials to new hardware,
and students may theoretically continue to interact with material immediately by applying
it to their own questions.

\subsubsection{Challenges of a traditional bootcamp approach}

The primary drawback of this approach comes in the form of variability and
reliability. Because students bring their own computing environments with them,
the experience with the course depends heavily on whether materials were
successfully able to run on their laptops. For example, some students have
difficulty setting up packages and environments, this costs time at the
beginning of class and often creates interruptions throughout the day. In
addition, some have pre-existing installations of software that may clash with
new installations, or introduce version dependency conflicts. Finally, because
computational power varies across students, it limits the kind of tutorials that
can be given by the instructor. This variability also has
implications for equality and discrimination, as students without access to
laptops with the proper hardware for completing the course are often frustrated
and unable to learn as effectively \cite{clark-proc-scipy-2014}.

Another challenge with this approach is that instructors have limited control
over the software experience of each student. For example, because operating
systems have different file system structures, there may be broken paths or
incorrect function calls in the course material. In addition, because students
will pull the material for the day onto their own computers, the material must
be frozen relatively early. It is complicated for instructors to update
materials during the workshop unless they want to lead the class through a
session on pulling from Github.

Course materials that are developed and tested only on the instructors' laptops
may have software dependencies or other assumptions that are exposed too late or
during the workshop itself, leading to complications on the students'
laptops. This may be further compounded by differences in laptop and software
configuration across multiple instructors with mutually incompatible
requirements. Though learning how to deal with these conflicts is a valuable
skill to learn, teaching these skills is generally not the primary goal of
the course.

\subsubsection{Benefits of the cloud-computing approach}

The key benefit of using a cloud-computing approach is that it allows one to
easily standardize the experience of each student via a shared online
resource. Because computing environments are initialized automatically with
Docker and a cloud provider (in our case, the XSEDE Jetstream cloud),
students incur virtually zero startup cost
before interacting with the course materials. They only need a web browser in
order to run Jupyter notebooks that are hosted remotely.
This also ensures that all students
start with a ``clean'' computing environment - they have access only to the data,
scripts, and packages that were required for the class. As one of the workshop
participants explained, ``Hosting
the tutorial software made it seamless for users. I've never seen a hands on
tutorial work that well.''

This approach afforded many benefits for instructors and course development.
By utilizing Docker for student deployments, the instructors knew the
computational resources that would be available to the students. As such, they
could scale the demands of their scripts accordingly, such that everything would
``just work'' once dozens of students simultaneously attempted to run their
code. Because each student's computing environment was provisioned in its own VM
instance in the cloud, this ensured that each student would receive the full
computational power available to them. This is in contrast to a traditional shared server
environment the students all share the resources of a single machine and may
interfere with each other if they launch demanding computations at the same
time.

By hosting student materials in the cloud, this approach also allowed instructors
to access these computing environments during the class itself. Any updates,
changes to data, or new scripts could be pushed silently via the XSEDE Jetstream
platform, which reduced the difficulties associated with asking students to
download new data. This all served to allow instructors and students to focus on
the primary goal of the event: covering course material and getting students
familiar with the domain-specific analysis.

An extra benefit of this approach is portability, as individuals have the
ability to perfectly replicate the course on any new computing environment. With
a few Docker commands, it is possible to recreate the user experience on another
cloud-computing environment or even a person's laptop, provided that they had
the hardware capacity to handle course computations and a minimal understanding
of the Docker environment. This adds to the reproducibility of the course, and
lowers the barrier to entry for users to discover and interact with the
materials in the future. It also makes it easier for instructors to build new
materials based on the Docker images of the course, which encourages
collaboration and reduces the tendency for instructors to repeat efforts.

Finally, it is worth emphasizing that while this bootcamp used
the XSEDE cloud platform for managing computational resources (due to its
presence at many universities in the United States),
one could utilize any computing platform that provides full root access for user instances
running Docker containers. By utilizing a high-level cloud service (such as XSEDE Jetstream) built on
top of lower-level cloud computing technologies (such as OpenStack), the
teaching team required
minimal setup and expertise in connecting course materials with the resources in the cloud.

\subsubsection{Challenges of the cloud-computing approach and areas for improvement}

While using a cloud-computing approach for bootcamp-style pedagogy provided many
benefits, it also uncovered new challenges in development and execution of the
course. This section covers a few specific topics that could be addressed in
future iterations of this approach towards teaching.

\begin{enumerate}

\item {\bf Team knowledge in cloud-computing infrastructure}: The primary drawback of
this approach is the necessity of at least one team member to have knowledge
in computing infrastructures. While there are many available resources to gain an
understanding of cloud-computing infrastructure, they are often idiosyncratic and
inaccessible for novice users. Fortunately, many university campuses
have individuals who are trained in these methods, or have access to experts in
the community available through the Campus Champion program\footnote{\url{https://www.xsede.org/web/campus-champions}}.
In our case, instructors worked with our local Campus Champion who had
cyberinfrastructure expertise in order to set up the Docker images within
the XSEDE Jetstream environment. In addition, significant development had to go
into creating and debugging Docker images, particularly early on in the
development cycle of the class. In the future, it will be important to provide
practical guides for accessing and interfacing with campus-supported cloud computing infrastructure, as
well as a well-explained workflow for how to leverage these resources in
teaching. There is also opportunity for advances in software to mitigate this
problem, such as using the JupyterHub \cite{perez2015project} platform
for managing user instances.

\item {\bf Expenses associated with cloud computing}: Another challenge of using
commercial cloud computing is that computation time incurs a cost, either
financial or in the form of an allocation of Service Units (SUs).
Fortunately, the XSEDE allocation process has a quick
turnaround for Jetstream Education Allocations with a limit up to 50,000 SUs.
It is also common to receive allocations in the form of grants or free credits
from commercial cloud providers for educational use, though the process is often
opaque and not guaranteed.

In the absence of freely-avialable computational resources, it is also
possible to utilize any cloud platform's standard payment model in order
to pay for course infrastructure. For our course, the equivalent cost
for our setup on a commercial cloud provider would be approximately \$600-700.
This cost will likely go down in the future as infrastructure for more efficiently
managing cloud resources is created (see \cite{perez2015project}).

For situations where instructors do not have access to free credits for
commercial cloud computing or funding to pay for their own
cloud resources, universities should provide modest grants to pay for
these resources or provide Jetstream-like cloud computing resources at the
campus-level.

\item {\bf Knowledge in container technology}: While Docker is extremely powerful,
it is still growing in both its features and API. This can be an impediment to
sustainable developments based on Docker, and a challenge to novices that are
learning how to deploy course materials for the first time. In order to improve
the accessibility of this approach for new instructors, it will be crucial to
minimize the effort required to set up a minimal computational environment with
Docker. Fortunately the Jupyter project's \textit{datascience-notebook}\footnote{\url{https://github.com/jupyter/docker-stacks/tree/master/datascience-notebook}}
Docker image provides a good starting point, though more efforts towards streamlining
this process are needed.

\item {\bf Migrating students from the cloud to their computers}: A final
challenge is that students are working on a new and
unfamiliar computational environment, rather than on their own laptops. Because
all course materials are run in the cloud, a migration step is
needed to transfer the software and data to other, more familiar environments.
It is straightforward to initialize the course Docker image on a
personal laptop with a
few simple commands, but this can be a large impediment for a student who has
just learned how to program.

We recommend including guides specifically for
students who wish to migrate their work onto their laptops, or holding special
sessions to ``off-board'' students after the class is finished. This serves as a
natural parallel to the ``install-fest'' that often happens at the
beginning of traditional bootcamps. However, in this case the
off-boarding process occurs after significant learning and experience in
software and computing has been gained, potentially mitigating the frustration
associated with migrating off of the cloud. Meanwhile, the experience also
serves to familiarize students with the cloud-computing environment. As a result
they may be poised to do their own work on cloud-computing environments
provided by their university or through XSEDE (a practice that is increasingly common).

\end{enumerate}

\subsection{Next steps for improving the cloud-based bootcamp}

With these considerations in mind, we recommend the following tangible
advancements in order to make this type of cloud-based bootcamp as effective
as possible:

\begin{enumerate}

\item {\bf Technological improvements}: The largest technological hurdle we
faced was in streamlining the connection between a Docker image and a
cloud-based system that can deploy many instances of this image automatically.
As standards in cloud-computing platforms solidify, this approach would
benefit from a software platform that requires only a Docker image (even better,
only a Github repository) as well as the credentials for a cloud-based system
that supports a common platform such as Kubernetes. The platform could then
support giving a list of usernames / email addresses, and would automatically
create instances of the learning environment along with their corresponding IP
addresses.

There are several software efforts taking place along these
lines, including the JupyterHub \cite{perez2015project} and Binder\footnote{\url{http://mybinder.org}}
projects. These both use Kubernetes Helm Charts\footnote{\url{https://github.com/kubernetes/charts}} to
facilitate the deployment of interactive computing environments.
We hope that in the future the Jetstream Service Provider will provide
formal support for Kubernetes, as its underlying system is based on OpenStack,
which provides Kubernetes support in recent releases.
However, it should be possible for Jetstream end users to create individual VMs
and install Kubernetes themselves to take advantage of the Helm Chart for JupyterHub.
For instructions on setting up JupyterHub to run on cloud providers that provide
Kubernetes support, see \textit{Zero to Jupyterhub}
\footnote{\url{https://zero-to-jupyterhub-with-kubernetes.readthedocs.io}},
a guide developed recently by the Jupyter team in collaboration with the Berkeley campus.

\item {\bf Instructional improvements}: Regardless of any changes to hardware
or software, a collection of instructional materials should be created that
focuses on instructors without the technological know-how of a systems administrator.
Docker is complicated, but it can be effectively used by following simple,
straightforward guides. In addition, cloud resources such as XSEDE need to
improve their instructional materials in order to
streamline the onboarding process.

\item {\bf Cost and payment opportunities}: While running student instances in the cloud
is not a large amount of money for most class sizes, the
cost of deploying a cloud-based bootcamp on commercial clouds is still a
significant barrier to many instructors. We recommend that universities provide
a fast-track process of small grants or allocations with quick turnarounds
specifically to provide access to computing time on cloud-based infrastructure
that is pre-configured with container support for the purpose of teaching
these kinds of container-based bootcamps.

\subsection{Summary and future work}

This project represents a first step towards a flexible and easy way to deploy
computational environments on cloud platforms for the purposes of teaching
data analytic methods to scientists. These cloud platforms are generally available
through national advanced cyberinfrastructure (ACI) such as XSEDE, as well as on several
commercial providers, making it easy to deploy these technologies for
bootcamp-based pedagogy at many universities around the country.

It should be noted that this is a rapidly growing set of tools and practices,
and the landscape of what is possible will likely change quickly. For example,
the JupyterHub project \cite{perez2015project} is being developed in parallel with
these efforts, and will streamline the process of connecting user
accounts with cloud providers. In addition, the best-practices around using
these tools effectively for pedagogy will continue to improve as the community
gains a better understanding of how to leverage this cloud-based approach.

\end{enumerate}

\section{Methods and technical infrastructure}\label{sec:methods}

This section describes the cyber-infrastructure support that made this workshop possible, and
highlights reusable and shareable patterns to build on for future work. We will
begin with a rough guide to the steps required to replicate this style of
bootcamp elsewhere, and then cover more details as to the specific hardware and
software implementations.

For instructors interested in hosting a similar event at their institution, the
following steps should be taken:

\begin{enumerate}
\item Choose a cloud provider (examples using XSEDE are inset below)
  \begin{enumerate}
    \item Contact your local XSEDE Campus Champion
    \item Decide on the resources needed (e.g. number of nodes, GPUs, etc)
    \item Make a plan for spin-down of cloud instances after the class
    \item Apply for an XSEDE Education Allocation for Jetstream
  \end{enumerate}

\item Install Docker for Mac or Docker for Windows on a laptop
\item Build docker container based on Jupyter's \emph{datascience-notebook}\footnote{\url{https://github.com/jupyter/docker-stacks/tree/master/datascience-notebook}}
\item Instructors create course materials on a shared github repo
\item Test instructor's Jupyter notebooks in Docker on laptop
\item Deploy user containers
  \begin{enumerate}
  \item Provision virtual machines (VMs) on Jetstream
  \item Deploy custom docker container on Jetstream VMs
  \end{enumerate}
\item Send IP addresses for each machine to students (one per student)
\item Connect to Jupyter notebooks running on Jetstream via a public IP address
\item Run the bootcamp
\item Assist students in migrating work onto their own machines
\end{enumerate}

All Jupyter notebook content for the course is available at:\\
\indent\indent\url{https://github.com/choldgraf/UCSF-Data_Driven_Neuro}

All build and deploy scripts are available at:\\
\indent\indent\url{https://github.com/aculich/UCSF-Data_Driven_Neuro-deploy}

\begin{figure}[h]
\centering
\includegraphics[width=0.5\textwidth]{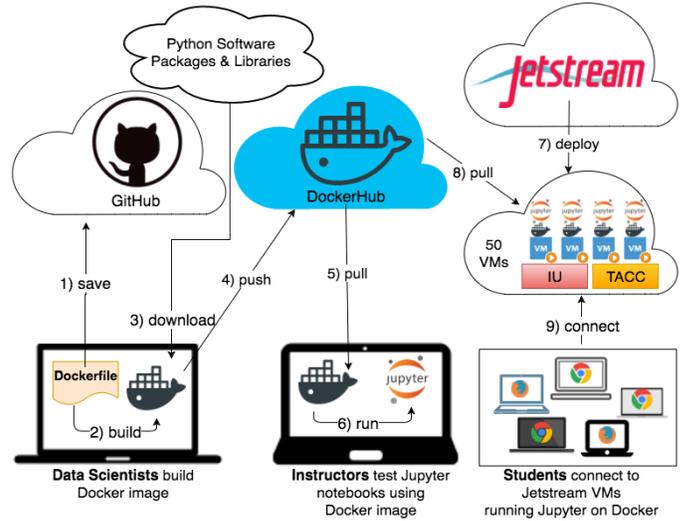}
\caption{Overview building Docker container and deploying to Jetstream for bootcamp}
\end{figure}

\subsection{Computational resources and the XSEDE Jetstream cloud}

The Jetstream~\cite{Stewart2015Jetstream} cloud platform
\footnote{\url{http://jetstream-cloud.org}} provided the computational resources for the
workshop. Jetstream's core capabilities include the ability to create
interactive Virtual Machines (VMs), access to remote desktops through a web
browser, and publishing VMs with a Digital Object Identifier (DOI). Jetstream is
attractive because it provides researchers a simple web-based
interface\cite{NiravCyberinfra2016} to launch, provision, manage, build, and
share customized virtual machines that include complete software dependencies
for running complex applications, whereas HPC environments traditionally do not
provide full administrative (root) access and are often not as flexible as
cloud-based virtual machines.

\begin{figure}[h]
\centering
\includegraphics[width=0.5\textwidth]{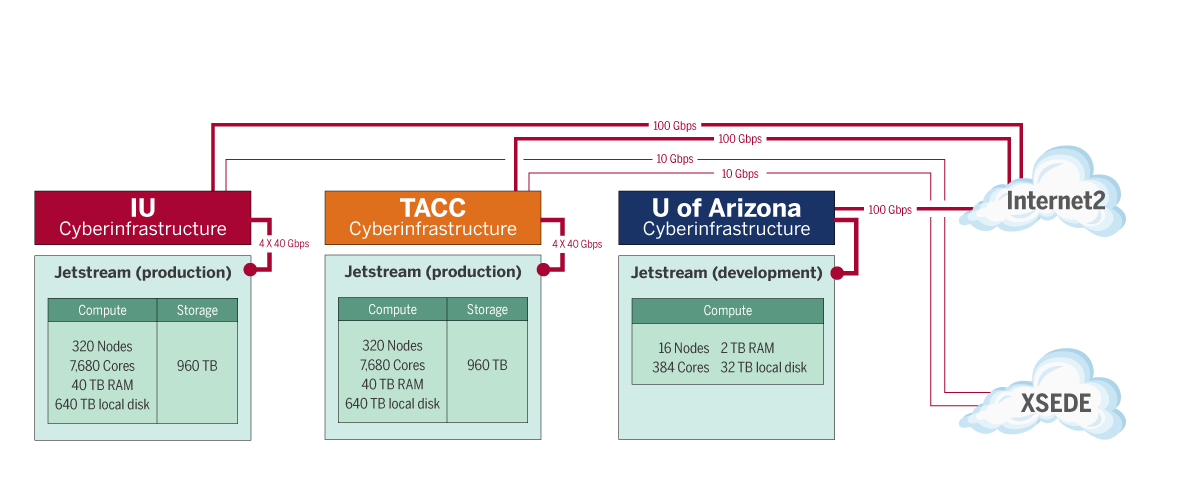}
\caption{XSEDE Jetstream cloud production infrastructure provided by Indiana University (IU) and Texas Advanced Computing Center (TACC). Taken from \url{http://jetstream-cloud.org/technology.php}}
\end{figure}

Access to Jetstream is available to researchers at no cost through the
NSF-funded XSEDE\footnote{\url{https://www.xsede.org/}} (Extreme Science and
Engineering Discovery Environment) project~\cite{Towns2014XSEDE}. This
offers access to a plethora of supercomputers, cloud environments, as well as high-end
visualization and data-analysis resources across the
country in order to address increasingly diverse scientific and
engineering challenges.

To obtain access, a qualified Principal Investigator writes a resource
justification and submits an allocation request. To help speed up the
process of choosing and obtaining access to the resource, many campuses
have local XSEDE Campus Champions who can facilitate quick access and
help prepare an allocation request.

For the neuroimaging workshop, the local XSEDE Campus Champion worked with Berkeley
Institute for Data Science (BIDS) and eScience Institute data scientists to
prepare an Education Allocation request. Below are some key excerpts from the
1-page allocation request\footnote{\url{https://portal.xsede.org/documents/10308/29438/Jetstream+Education+Allocation+request+-+Sample/28517ffe-79fa-4e3f-98c9-b64f126a1e6b}},
which you can read in full from the list of example allocation requests:

\begin{itemize}
\item 50 Virtual Machines running simultaneously \\(40 students + 5 instructors +
test/spare/debug VMs)
\item Each VM will need to be a: Jetstream m1.medium VM \\(6 vCPUs, 16GB RAM, 60GB
  Storage)
\item Each VM will need an external IP address so students can connect remotely
  with a web browser to a Jupyter Notebook running on the machine
\item We are requesting 10,000SUs in total.
\end{itemize}

An SU is a Service Unit. The maximum number of SUs for an Education Allocation
on Jetstream is 50,000SUs, however after we calculated the total resources we
needed for the course, we determined that 10,000SUs would be sufficient to
conduct the course, as well as allow students to run VMs for a short time
following the event. The SU cost per hour for each VM can be determined at the
Jetstream General Virtual Machine Configurations page\footnote{\url{http://jetstream-cloud.org/general-vms.php}}.
At the time of the workshop the m1.medium VM noted above cost 6 SUs per hour.

The technology we used to deploy the workshop in addition to the Jetstream cloud
platform includes Docker, Dockerhub, and the datascience-notebook docker-stacks\footnote{\url{https://github.com/jupyter/docker-stacks}}
maintained by the Jupyter project.

\subsection{Development and environment control with Docker}

Each of the instructors initially used their own laptops to develop Jupyter
Notebook-based tutorials on computer vision and machine learning for
neuroscience, using state-of-the-art deep learning methods.

Research IT staff worked with BIDS and eScience data scientists to build a
customized container from the Jupyter project's datascience-notebook image. This
provides a pre-configured Jupyter Notebook 4.3.x; Conda Python 3.x and Python
2.7.x environments; and several common libraries including: pandas
\cite{mckinney-proc-scipy-2010}, matplotlib \cite{hunter2007matplotlib}, scipy
\cite{scipy}, seaborn \cite{michael_waskom_2014_12710}, scikit-learn
\cite{Pedregosa2012-dm}, and scikit-image \cite{van2014scikit}. Additional
neuroscience-specific packages were included such as Dipy for diffusion magnetic
resonance imaging (dMRI) analysis \cite{Garyfallidis2014FrontNeuroinf}.

This customized container ensured that all the students had an identical
environment on the day of the workshop, including all required software
dependencies. The container made it possible for participants to easily run the
software without installing each of the components, often a lengthy and
error-prone process at the start of many workshops. The container can also be
used to tag versions of the environment, such that the software is preserved
for future use. Months or years from now it will be possible to re-run the
notebooks again, even if external software packages and dependencies have
changed.

The container image was pushed to Docker hub\footnote{\url{https://hub.docker.com}},
which provides a centralized resource for container image discovery,
distribution and change management, user and team collaboration, and workflow
automation. Once a Docker image is on Docker hub, it can be downloaded and run
with a single command. At a high-level the process is:

\begin{enumerate}
  \item On a laptop, create a Dockerfile that:
  \begin{itemize}
    \item derives \texttt{FROM jupyter/datascience-notebook}
    \item installs workshop-specific software packages
    \item pin packages with explicit versions defined
  \end{itemize}
  \item Build docker image
  \item Run docker image as container
  \item Test Jupyter notebooks running inside container
  \item Push docker image to DockerHub
\end{enumerate}

\subsection{Putting it together}

On the day of the workshop, the 50 Jetstream virtual machines (VMs) were
deployed by hand using Jetstream's Atmosphere \cite{NiravCyberinfra2016} web interface\footnote{\url{http://www.cyverse.org/atmosphere}}. While it is possible to create
scripts for a fully automated deployment using the low-level OpenStack API that
Jetstream is built on, we decided that the additional complexity was not
desirable. Using the Atmosphere web interface is a quick and simple 6-step
process that allowed us to manually start the deployment of all 50 instances in
just a few minutes. At a high-level the process is:

\begin{enumerate}
\item Select the pre-defined VM image: \\Ubuntu 14.04.3 Development GUI
\item Choose instance size: m1.medium VM \\(6 vCPUs, 16GB RAM, 60GB
  Storage)
\item Click ``Advanced Options''
\item Select deployment script from github URL for the workshop
\item Click ``Continue to Launch''
\item Click ``Launch Instance''
\end{enumerate}

\begin{figure}[h]
\centering
\includegraphics[width=0.5\textwidth]{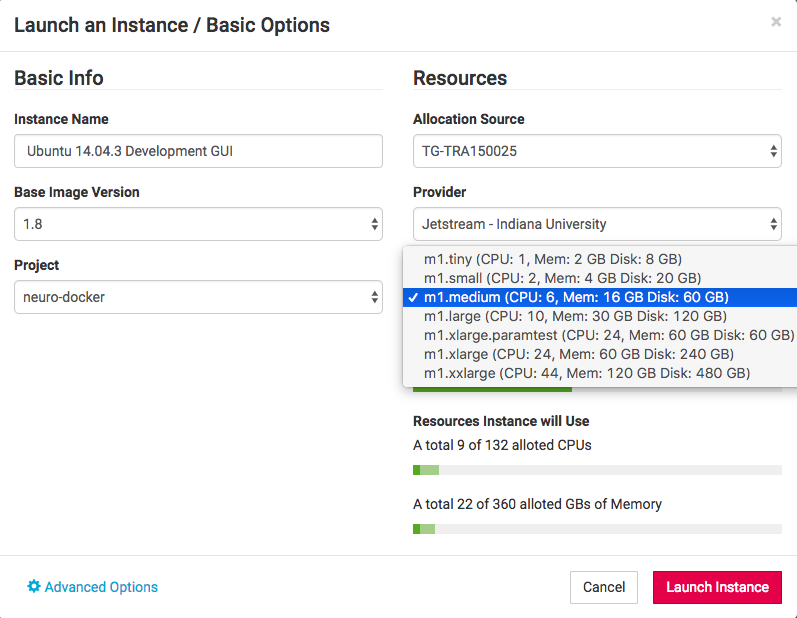}
\caption{Screenshot of configuring a virtual machine (VM) to launch on Jetstream}
\end{figure}

The deployment script from github URL for the workshop is a simple bash script
that will be run when the VM starts. It does the following:

\begin{enumerate}
\item Install Docker on the Jetstream VM
\item Pull the workshop docker image from DockerHub
\item Download all the data needed for the workshop examples
\item start the Docker container running the Jupyter notebook, password-protected on a standard web port (80/443)
\end{enumerate}

After the workshop the participants were allowed to continue accessing their
notebook on the Jetstream platform for a limited time using the Education
Allocation for the workshop. After the allocation expired, each individual could
either:

\begin{itemize}

\item install Docker for Mac or Docker for Windows to download and run the
      container on their own laptop
\item apply for their own Startup and Research Allocations on XSEDE Jetstream

\end{itemize}

\subsubsection{Considerations for security and privacy}

It is worth noting a few issues related to networking and security that must be
addressed for any scenario involving remote computing (whether in a cloud
computing environment or traditional server environment).

When running Jupyter Notebooks on a laptop, they typically listen on network
port 8888 and a user connects via their web browser to
\url{http://localhost:8888}. In this configuration, the port is not accessible
to a remote attacker. In a server environment, remote access is a key
feature, so it requires running the Jupyter notebook in a secure mode requiring
a token or a password. Thankfully, the Jupyter team has configured the
docker-stacks to run in a secure mode by default.

For the workshop we chose a single password to deploy to each of the Jupyter VM
instances and wrote the password on the whiteboard for participants. We also
copied the IP address of each machine into a Google spreadsheet, and assigned it
to each of the participants. Alternative solutions to this include using a link
shortening service such as \url{http://bit.ly} to generate a URL out of student
names.

\section{Conclusions}

The purpose of this manuscript is to provide inspiration
and a guide for how to utilize cloud-based technologies towards teaching a
bootcamp-style event. It also aims to lay out a path towards refining this
process to accommodate new research domains and training events, and to make it
more straightforward for instructors to set up course infrastructure without the
need for exceptional technical knowledge. Utilizing cloud-computing
infrastructure has the ability to improve both the teaching and learning
experience in data-heavy fields, and offers new opportunities for giving
researchers a pragmatic, hands-on experience with data that focuses on the
topics covered in the course. As the materials available for instructors
improves, we believe that this approach will increase in efficacy and become a
common approach in modern-day pedagogy.

\begin{acks}

This work was supported through a grant to the Berkeley Institute for Data
Science and the University of Washington eScience Institute from the Gordon \&
Betty Moore Foundation and the Alfred P. Sloan Foundation.

This work used the Extreme Science and Engineering Discovery Environment
(XSEDE), which is supported by National Science Foundation grant number
ACI-1053575.

The authors would like to thank Prof. Lea Grinberg for her support in organizing
the course described herein. We also wish to thank Aya Rokem for providing
a welcome early submission deadline for preparing course materials.

\end{acks}

\bibliographystyle{ACM-Reference-Format}
\bibliography{paper}

\end{document}